\documentstyle[amsfonts,aps]{revtex}
\begin{document}
\draft
\tighten
\twocolumn[\hsize\textwidth\columnwidth\hsize\csname   
@twocolumnfalse\endcsname                             
\title{ Diffusion-controlled annihilation $A + B \rightarrow 0$
with initially separated reactants: The death of an $A$ particle island in
the $B$ particle sea}
\author{ Boris ~M.~Shipilevsky}
\address{ Institute of Solid State Physics, Chernogolovka,
Moscow district, 142432, Russia}
\date{\today}
\maketitle
\begin{abstract}
We consider the diffusion-controlled annihilation dynamics $A+B\to 0$ 
with equal species diffusivities in the system where an island of particles
$A$ is surrounded by the uniform sea of particles $B$. We show that once the
initial number of particles in the island is large enough, then at any
system's dimensionality $d$ the death of the majority of particles occurs in
the {\it universal scaling regime} within which $\approx 4/5$ of the
particles die at the island expansion stage and the remaining $\approx 1/5$
at the stage of its subsequent contraction. In the quasistatic approximation
the scaling of the reaction zone has been obtained for the cases of
mean-field ($d \geq d_{c}$) and fluctuation ($d < d_{c}$) dynamics of the
front.

\end{abstract}
]                                                 

\narrowtext

The reaction front formed in the $A + B \rightarrow 0$
reaction-diffusion systems with initially separated reactants is of great
interest since it represents a pattern for a wide spectrum of processes in
physics, chemistry, and biology [1]. Since the seminal article of
Galfi and Racz [2] much work has been devoted to studying this problem
by different approaches[3-20]. A standard way to treat the problem
analytically is to solve the system of equations
\begin{eqnarray}
\partial a/\partial t = D_{A}\nabla^{2} a - R, \quad
\partial b/\partial t = D_{B}\nabla^{2} b - R
\end{eqnarray}
with the initial state given by
\begin{eqnarray}
a(x,0)=a_{0}\theta(-x), \quad b(x,0)=b_{0}\theta(x),
\end{eqnarray}
where $a(x,t)$ and $b(x,t)$ are the mean local concentrations of $A's$ and
$B's$, $R(x,t)$ is the macroscopic reaction rate and $\theta(x)$ denotes the
Heaviside step function, so that the $A's$ are initially uniformly
distributed on the left side ($x< 0$), and the $B's$ on the
right side ($x> 0$) of the initial boundary.

Dimensional [9,14] and renormalization group [11,12] analyses show that
at $d \geq d_{c}=2$ one can adopt the mean-field approximation
$R(x,t)=ka(x,t)b(x,t)$($k$ being the reaction rate constant) with logarithmic
corrections in the $2d$ case, whereas in $1d$ systems
fluctuations play the dominant role and the explicit form of $R$ remains
unknown. There are, however, several techniques which enable one to derive
a lot of information from (1) even for $d < d_{c}$. They are focused on
the long-time limit $kt\to\infty$ and include, as two basic concepts,
the scaling ansatz [2] and the quasistatic approximation [8,9,11].
According to the scaling ansatz (SA) the long-time behavior of the system
inside the reaction zone may be represented in the form
\begin{eqnarray}
R=R_{f}{\cal Q}\left(\frac{x-x_{f}}{w}\right),
\end{eqnarray}
where $x_{f}\propto t^{1/2}$ denotes the position of the reaction
zone center, $R_{f}\propto t^{-\beta}$ is the height, and
$w\propto t^{\alpha}$ is the width of the reaction zone. At $d \geq d_{c}$
the scaling exponents take the values $\alpha=1/6, \beta=2/3$[2] whereas at
$d=1$ they appear equal to $\alpha=1/4, \beta=3/4$[9-14], so that at any
$d$ the width of the reaction zone on the diffusion length scale
$\Lambda_{D} \propto t^{1/2}$ asymptotically unlimitedly contracts:
$$
w/\Lambda_{D} \to 0 \quad {\rm as} \quad t \to \infty.
$$
The quasistatic approximation (QSA) consists in the assumption that for
sufficiently long times the kinetics of the front is governed by two
characteristic time scales. One time scale, $t_{J}=-(d\ln J/dt)^{-1}$,
controls the rate of change in the diffusive current $J=J_{A}=J_{B}$ of
particles arriving at the reaction zone. The second time scale,
$t_{f}\propto w^{2}/D$ is the equilibration time of the
reaction front. Assuming $t_{f}/t_{J}\ll 1$ from the QSA in the mean-field
case with $D_{A,B}=D$ it follows [8,9]

\begin{eqnarray}
R_{f}\sim J/w, \quad
w\sim (D^{2}/Jk)^{1/3},
\end{eqnarray}
whereas in the $1d$ case $w$ acquires the $k$-independent form
$w\sim (D/J)^{1/2}$ [9,11,17]. The most important feature of the QSA is that
$w$ and $R_{f}$ depend on $t$ only through the time
dependent boundary current $J(t)$, which can be calculated analytically
without knowing the concrete form of ${\cal Q}$, i.e., in fact, representing
the reaction zone on the scale $\Lambda_{D}$ in the form
$R(x,t)=J\delta(x-x_{f})$. On the basis of the QSA the general
description of spatiotemporal behavior of the system $A+B\to 0$ has been
obtained for arbitrary nonzero diffusion coefficients and initial species
concentrations [19]. These results are in full agreement with extended
numerical calculations and experiments and were generalized recently to
the cases of reversible reaction $A+B \leftrightarrow C$ [21-24] and to
several more complex reactions [25-29].

Until now, however, the main attention has been focused on the systems with
$A$ and $B$ domains unlimited extension, i.e., with unlimited number of
$A's$ and $B's$ particles. The aim of this work is to develop
{\it a new line} in the study of the $A+B\to 0$ dynamics under the
assumption that the particle number of one of the species, say $A$, per unit
of the initial boundary is finite. More precisely, we will consider the
problem on the dynamics of death of an $A$ particle island surrounded by
the uniform sea of particles $B$ and will reveal the defining features
of this process.

Let particles $A$ with concentration $a_{0}$ be uniformly distributed
in the island $x\in (-L,L)$ surrounded by the unlimited sea of particles $B$
with concentration $b_{0}$ on the left $x\in (-\infty, -L)$ and
on the right $x\in (L, \infty)$ of the island. By symmetry our problem
is reduced to the solution of the system (1) in the interval
$x\in [0, \infty)$ at the initial conditions
$$
a(x,0)=a_{0}\theta(L-x), \quad  b(x,0)=b_{0}\theta(x-L)
$$
with the boundary conditions
$$
\nabla (a, b)\mid_{x=0}=0, \quad b(\infty, t)=b_{0}.
$$
To simplify the problem essentially we will assume, as usually,
$D_{A}=D_{B}=D$. Then by measuring the length, time, and concentration in
units of $L$, $L^{2}/D$, and $b_{0}$, respectively,
i.e. assuming $L=D=b_{0}=1$, and defining the difference concentration
$s(x,t)=a(x,t)-b(x,t)$ and the ratio of initial concentrations $a_{0}=r$
we come from (1) to the simple diffusion equation
\begin{eqnarray}
\partial s/\partial t = \nabla ^{2} s
\end{eqnarray}
with the initial conditions
$$
s_{0}(x\in [0,1))=r, \quad s_{0}(x\in (1,\infty))=-1,
$$
and the boundary conditions
$$
\nabla s\mid_{x=0}=0 , \quad s(\infty,t) = -1.
$$
The solution to Eq.(5) has the form
\begin{eqnarray}
s(x,t)=\frac {r+1}{2}\left[{\rm erf}\left(\frac{1+x}{2\sqrt{t}}\right) +
{\rm erf}\left(\frac{1-x}{2\sqrt{t}}\right)\right]-1,
\end{eqnarray}
whence, according to condition $s(x_{f},t)=0$ [2,19] there immediately
follows the equation defining the law of motion of the reaction front
center, $x_{f}(t)$
\begin{eqnarray}
{\rm erf}\left(\frac{1+x_{f}}{2\sqrt{t}}\right) +
{\rm erf}\left(\frac{1-x_{f}}{2\sqrt{t}}\right)= \frac{2}{r+1}.
\end{eqnarray}
Let us assume that $d \geq d_{c}$ and the reaction rate constant $k$ is
sufficiently large [30] so that at times $t\sim t_{GR}\propto k^{-1} \ll 1$
the annihilation goes to the scaling Galfi-Racz regime, i.e. in the
vicinity of $x_{f}$ there forms a narrow reaction zone $w/\Lambda_{D}\ll 1$.
At $t \ll 1$ and $|1-x_{f}| \ll 1$ from (7) in accordance with [2] we find
$$
x_{f}=1+c_{f}\sqrt{t}+\cdots,
$$
where ${\rm erf}(c_{f}/2)=(r-1)/(r+1)$ and, hence, at $r\leq 1$ the island
contracts, whereas at $r > 1$ the island expands. By virtue of the fact
that the number of particles in the island (per unit of the initial boundary)
is finite, the stage of its expansion always goes to the stage of its
subsequent contraction to end in an instant of time $t_{c}$, when the
reaction front center approaches the origin of coordinates,
$x_{f}(t_{c})=0$:
\begin{eqnarray}
{\rm erf}(1/2\sqrt{t_{c}})=1/(r+1).
\end{eqnarray}
According to (8) at $r \gg 1$ the "lifetime" of the island $t_{c} \gg 1$
so the majority of the particles die at times $t \gg 1$, when the
diffusive length exceeds appreciably the initial island size,
$\Lambda_{D} \gg 1$. The evolution of the island in such a large-$t$ regime
is of principal interest to us here, and its analysis is the main goal of
the present report.

In the limit $r,t \gg 1$ from (6) we find
\begin{eqnarray}
s(x,t)=\frac{(r+1)}{\sqrt{\pi t}}e^{-x^{2}/4t}(1-\chi)-1,
\end{eqnarray}
where $\chi=(1-x^2/2t)/12t +\cdots$, and, hence, the law of the front
motion is
\begin{eqnarray}
x_{f}=2\sqrt{t}(1+\epsilon)
\ln^{1/2}[\left(\frac{r+1}{\sqrt{\pi t}}\right)(1-\epsilon)],
\end{eqnarray}
where $\epsilon= 1/12t + \cdots$. By assuming $t \gg 1 > t_{GR}$, at the
stage of the island expansion and at a considerable time period of its
subsequent contraction, the duration of which will be estimated below, the
reaction zone width is $w \ll x_{f} < \Lambda_{D}$, therefore in terms of
the QSA we can take $w/x_{f}\to 0$, and, hence, $R(x,t)=J\delta (x-x_{f})$
so that $a=s, b=0$ when $x<x_{f}$, and $a=0, b=|s|$ when $x>x_{f}$. Thus, by
neglecting the width of the reaction zone, for the number of particles in the
island we have  $N(t)= \int_{0}^{x_{f}}a(x,t) dx=\int_{0}^{x_{f}}s(x,t) dx$
whence, with account taken of (9) and neglecting the contribution
of $\chi$, we find
\begin{eqnarray}
N(t)=(r+1){\rm erf}(x_{f}/2\sqrt{t})-x_{f}.
\end{eqnarray}

Let us now calculate the maximal amplitude $x^{M}_{f}$ of the island
expansion and the time $t_{M}$ of its achievement. According to (7) for
$r>1$ from condition $\dot{x}_{f}=0$ we have exactly
$$
x^{M}_{f}=2t_{M}{\rm Arcoth}(x^{M}_{f})
$$
whence for $r \gg 1$ it follows
\begin{eqnarray}
x^{M}_{f}=\sqrt{2t_{M}}(1+1/12t_{M}+\cdots).
\end{eqnarray}
Neglecting the terms $O(1/r^{2})$ from (10),(12) we find
$$
t_{M}=(r+1)^{2}/\pi e, \quad
x^{M}_{f}=(r+1)\sqrt{2/\pi e}.
$$
Comparing $t_{M}$ with the lifetime of the island (8),
$$
t_{c}=(r+1)^{2}/\pi,
$$
we conclude that independently of $r$ the ratio $t_{c}/t_{M}=e={\rm const}$.
Going with allowance for this to the reduced coordinate
$\zeta=x/x^{M}_{f}$ and time $\tau=t/t_{c}$ we, finally, come to scaling
relations for the distribution of particles $s(\zeta,\tau)$
\begin{eqnarray}
s(\zeta,\tau)=e^{-\zeta^{2}/2e\tau}/\sqrt{\tau}-1,
\end{eqnarray}
for the center of the reaction front
\begin{eqnarray}
\zeta_{f}=\sqrt{e\tau|\ln\tau|},
\end{eqnarray}
and for the number of particles in the island
\begin{eqnarray}
N/N_{0}=\gamma_{r}{\cal G}(\tau),
\end{eqnarray}
where $\gamma_{r}=(r+1)/r\to 1$ as $r\to \infty$ and scaling function
$$
{\cal G}(\tau)= {\rm erf}(\sqrt{|\ln\tau|/2})- \sqrt{2\tau|\ln\tau|/\pi}.
$$
From (13) and (15) we find that in the turning point $\tau_{M}=1/e$
$$
a(0,\tau_{M}) = \sqrt{e}-1 = 0.64872...,
$$
$$
\gamma^{-1}_{r}N_{M}/N_{0}= 0.19886...,
$$
and, hence, independently of the initial number of particles, $N_{0}=r$,
$\approx 4/5$ of the particles die at the stage of the island expansion and
the remaining $\approx 1/5$ at the stage of its subsequent contraction. In
Figs.1 and 2 are shown the calculated according to Eqs.(6),(7) plots of the
behavior of the particle distribution $|s(\zeta,\tau)|$, of the center of the
reaction front $\zeta_{f}(\tau)$, and the number of particles
$\gamma^{-1}_{r}N(\zeta_{f})/N_{0}$, which give the full insight into the
scaling regime of the death of the island (13)-(15).

Let us come now to the question of the scaling behavior of the reaction rate
$R(x,t)$ in the vicinity of the reaction front center $x_{f}$.
Assuming according to [8,9,19] that within the applicability of the
inequalities $t_{f}/t_{J} \ll 1$ and $w \ll x_{f}$ the reaction rate
$R(x,t)$ can be described in terms of the QSA by Eqs.(3),(4) to calculate
$R_{f}$ and $w$ it only remains for us to calculate the diffusive current
$J(t)$ of particles arriving at the reaction zone. According to (9),(11)
we have
$$
J=-\dot{N}=-\partial s/\partial x\mid_{x=x_{f}}= x_{f}/2t
$$
whence it follows
\begin{eqnarray}
J/J_{M} = \sqrt{|\ln \tau|/e\tau},
\end{eqnarray}
where $J_{M}=1/x^{M}_{f}=\sqrt{\pi e/2}/(r+1)$. Substituting (16)
into Eqs.(4) we come to the scaling of the reaction zone
\begin{eqnarray}
R_{f}\sim (J^{4}k)^{1/3}=
R^{M}_{f}\left(\frac{|\ln\tau|}{e\tau}\right)^{2/3}
\end{eqnarray}
\begin{eqnarray}
w\sim (Jk)^{-1/3}=
w_{M}\left(\frac{e\tau}{|\ln\tau|}\right)^{1/6},
\end{eqnarray}
where $R^{M}_{f}\sim (k/r^{4})^{1/3}$ and $w_{M}\sim (r/k)^{1/3}$ so that
$wR_{f}\sim R_{global}=\int_{0}^{\infty}R dx= J$.
From (18) it is seen that at the stage of the island contraction the
logarithmic term becomes dominant,
and as $\tau \to 1$, i.e., $\delta\tau = 1-\tau \to 0$, the width of the
reaction front $w$ diverges as $w \propto (\delta\tau)^{-1/6}$. Comparing
the characteristic times $t_{J}=-(d\ln J/dt)^{-1}\sim r^{2}\delta\tau$ and
$t_{f}\sim w^{2}\sim (r/k)^{2/3}(\delta\tau)^{-1/3}$ we have
$t_{f}/t_{J}\sim (r^{2}k)^{-2/3}(\delta\tau)^{-4/3}$ whence it follows
that the characteristic time at which the QSA is violated is
$$
\delta\tau_{Q}\propto 1/r\sqrt{k}.
$$
Comparing then $w$ and $x_{f}$ we obtain
$w/x_{f}\sim \sqrt {t_{f}/t_{J}}\sim (r^{2}k)^{-1/3}(\delta\tau)^{-2/3}$
whence it follows that the width of the reaction zone becomes comparable
with the island size at the times
\begin{eqnarray}
\delta\tau_{w}\sim \delta\tau_{Q}\propto 1/r\sqrt{k}.
\end{eqnarray}
From (19) it is seen that at sufficiently large values of $k \gg 1$
the characteristic times $\delta\tau_{w} \sim \delta\tau_{Q}\ll 1/r \ll 1$
and, hence, practically all the particles die in the scaling regime (15).

The analogous estimations in the case of the $k$-independent
fluctuation $1d$ regime with $w \sim 1/\sqrt{J}$ yield the expressions
(17),(18) with $R^{MF}_{f}\sim r^{-3/2}, w_{M}^{F}\sim \sqrt{r}$, and
exponents $3/4$ instead of $2/3$, and $1/4$ instead of $1/6$, respectively.
In the end for the $1d$ case we have
\begin{eqnarray}
\delta\tau^{F}_{w}\sim \delta\tau^{F}_{Q}\propto 1/r^{2/3}.
\end{eqnarray}
According to (20), despite the much more "blurred" structure of the
$1d$ island, in this case, too, in the limit of sufficiently large
$r \gg 1$ the majority of the particles die in the scaling regime (15).

In summary, the problem of death of an $A$ particle island in
the $B$ particle sea at equal diffusivities of $A's$ and $B's$ particles has
been first considered. It has been found that at sufficiently
large initial number of particles in the island, $r \gg 1$, and sufficiently
large reaction rate constant, $k \gg 1$, the death of the
majority of the particles at any $d$ occurs in the universal scaling regime,
and the most essential features of this regime have been revealed.
The obtained results can have many applications, especially in surface
science, and, in particular, they can provide fresh insight into the dynamics
of the Ovchinnikov-Zeldovich hierarchic $A-B$ structures [1]. By analogy with
[19] it may be expected that the analysis presented can be easily extended
over the general case $D_{A}\neq D_{B}\neq 0$ (note that in the static case
$D_{B}=0$, which belongs to the separate universality class, with initially
$N_{0}$ particles $A$ at the origin of coordinates, in [16] the formally
coincident with (15) scaling, $N=N_{0}{\cal F}(t/N_{0}^{2})$, was found).
An investigation into the general case of nonzero diffusivities and a most
interesting generalization for the anisotropic diffusion of reactants is
expected to be presented in a future report.

I would like to thank Alexey Nekrasov for help in computer work.
This research was financially supported by the RFBR through Grant
No. 02-03-33122.

\begin{figure}
\caption {Evolution of the distribution of particles $|s(\zeta,\tau)|$
calculated according to Eqs.(6) and (7) for $r=10$. The left part of
the distribution ($\zeta < 0$) is obtained through the mirror reflection
of the right one ($\zeta > 0$). The region
$s > 0$ $(-\zeta_{f} < \zeta < \zeta_{f})$, belonging to the island, is
colored grey. For demonstration only a part of the picture is
shown for $\tau > 0.03$.}
\label{fig 1}
\end{figure}
\begin{figure}
\caption {(a) Inset: Time dependences $x_{f}(t)$ calculated from Eq.(7)
at $r=5$ (filled squares), $r=10$ (open squares), $r=20$ (filled circles),
and $r=40$ (open circles). Main panel: Collapse with the growing $r$ of the 
shown in the inset dependences to the scaling law (14)(solid line) in the 
rescaled coordinates $\zeta_{f}(\tau)$. For completeness is shown the motion 
of the both island fronts: $\pm x_{f}(t), \pm \zeta_{f}(\tau)$;
(b) Collapse with the growing $r$ of the calculated from Eqs.(6) and (7) 
dependences of the reduced particle number in the island 
$\gamma^{-1}_{r}N/N_{0}$ vs $\zeta_{f}$ to the scaling function 
${\cal G}(\zeta_{f})$(solid line): $r=5$ (filled squares), 
$r=10$ (open squares), $r=20$ (filled circles), and $r=40$ (open circles). 
The number of the island particles $N$ has been calculated by integrating 
Eq.(6) from $0$ to $x_{f}$ on assuming that $w/x_{f} \to 0$.}
\label{fig 2}
\end{figure}

\begin{references}
\bibitem{note} Note that at the chosen units of length, time, and
concentration the reaction rate constant $k$ is measured in units of
$D/L^{2}b_{0}$.
\end{references}
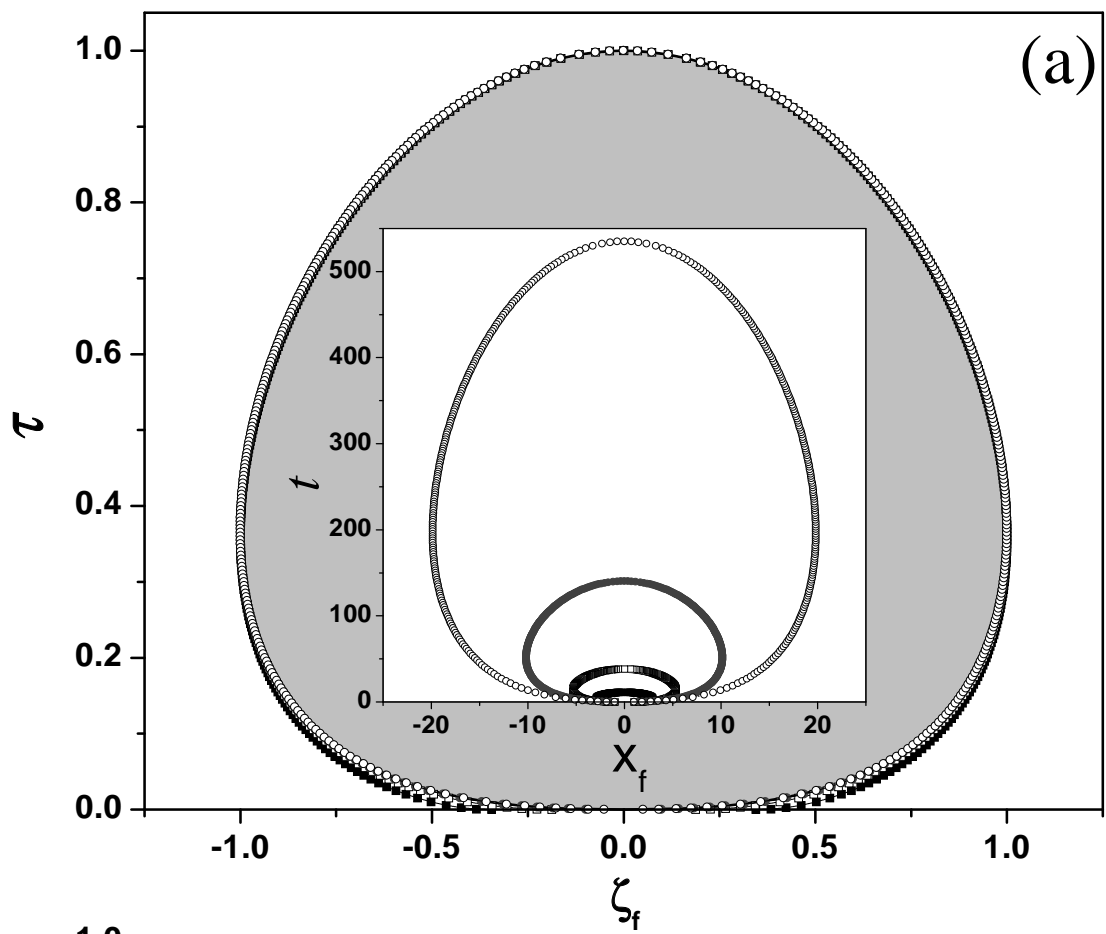
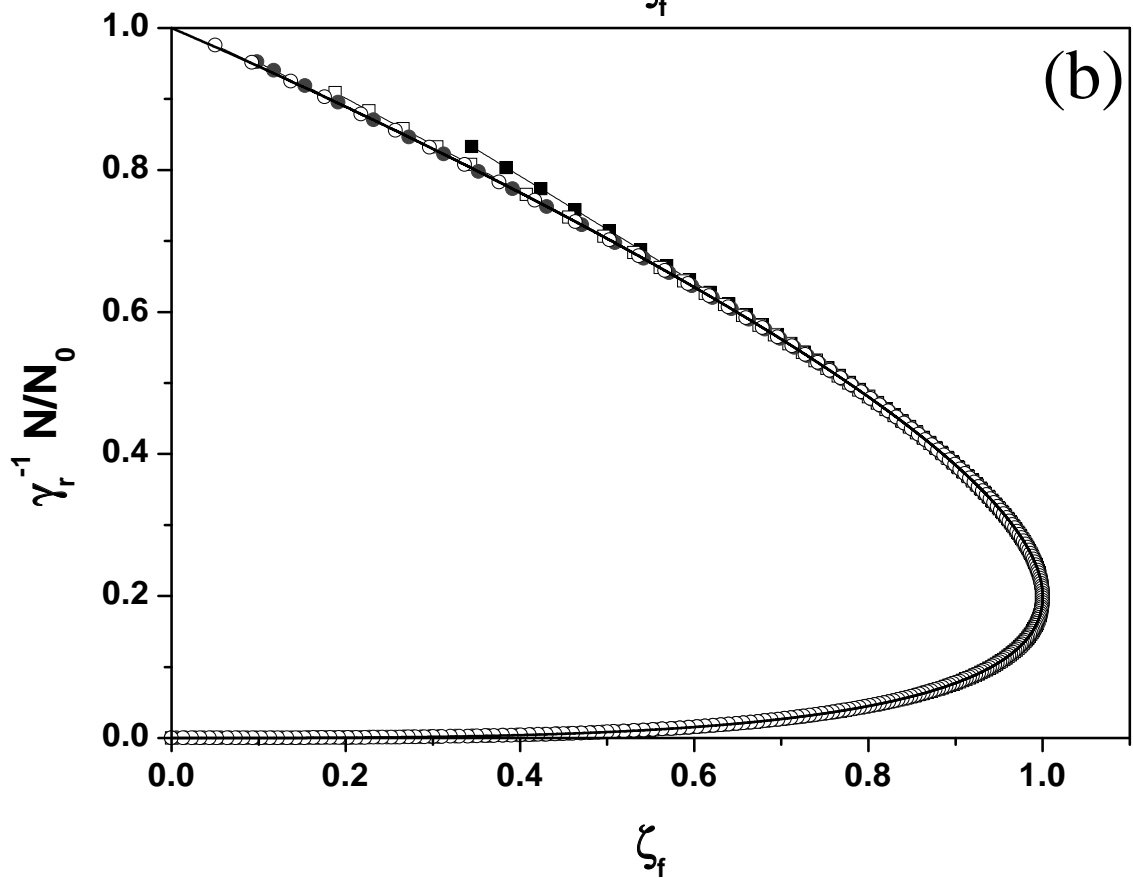
\end{document}